\documentclass[12pt]{article}
\usepackage{amsfonts}
\usepackage{amsthm}
\usepackage{amsmath}
\usepackage{amsfonts}
\usepackage{latexsym}
\usepackage{amssymb}
\usepackage[dvips]{color}
\DeclareMathOperator{\Real}{Re}
\DeclareMathOperator{\Imag}{Im}

\begin{document}

\title{The application Breit-Wigner form with radiative corrections to the resonance fitting}
\author{K. Yu. Todyshev  \thanks{e-mail:todyshev@inp.nsk.su}}
\date{}
\maketitle
{
\begin{center}
\noindent Budker Institute of Nuclear Physics, 630090,  Novosibirsk, Russia
\end{center}
}
\begin{abstract}
The nonrelativistic and relativistic Breit-Wigner 
forms are conventionally used for the resonance fitting. 
In this note we consider the application of the Breit-Wigner formula
with radiative corrections in initial state.
\end{abstract}

In this paper the problem of fitting a wide resonance  is considered.
The general expression for the cross section of $e^{+}e^{-}$
annihilation into hadrons via an onium resonance is
well known:
\begin{equation}
\sigma(W)=\frac{3\pi}{M^2}\int \!\,\frac{\Gamma_{ee}\Gamma_h}{(W\sqrt{1-x}-M)^2+\frac{\Gamma^2}{4}}\mathcal{F}(x,W) dx~,
\end{equation}
where W is the collision c.m. energy, 
$\Gamma$,~$\Gamma_{ee}$ and $\Gamma_h$ are total and partial
widths of the resonace,  M is its mass and $\mathcal{F}(x,W)$ 
is the probability to lose the fraction of energy x because of the
initial state radiation~\cite{bib:FADIN}. 
To obtain parameters of the resonances like $J/\psi$ and $\psi(2S)$ 
the additional convolution with a Gaussian distribution is usually  needed. 
Sometime it is possible to avoid double integration if simplify $\mathcal{F}(x,W)$ function.
The case of a narrow resonace is described in literature 
(see \cite{bib:JACKSON}). In Ref.~\cite{bib:ASIMOV} the calculation was made
for nonrelativistic Breit-Wigner in the soft-photon approximation,  in which
\begin{equation}
\mathcal{F}(x,W) = \beta x^{\beta-1}(1+\delta)~, 
\label{FadinKuraevSoftApp}       
\end{equation}
where $\beta=\frac{4\alpha}{\pi}(\ln\frac{W}{m_e}-\frac{1}{2})$,
$\delta=\frac{\alpha}{\pi}\Big(\frac{\pi^{2}}{3}-\frac{1}{2}\Big)+\frac{3}{4}\beta-\frac{\beta^2}{24}\bigg(\frac{2}{3}\ln\frac{W}{m_e}+2\pi^2-\frac{37}{4}\bigg)$~.\\

That would be usefull to estimate errors of the functions published in Ref.~\cite{bib:ASIMOV}, if make the similar calculations including the contribution of 
hard photons and the terms $\sim \frac{\Gamma}{W}$ .

Let us consider $\sigma(W)$ following~\cite{bib:ASIMOV} in the assumption that
$\frac{\Gamma}{W}\sim\beta$. We also include the terms $\sim \beta$ in  $\mathcal{F}(x,W)$,
which represent the contribution of hard photons in the first order by $\beta$.
With that accuracy, the $\mathcal{F}(x,W)$ is
\begin{equation}
\mathcal{F}(x,W) = \beta x^{\beta-1}(1+\delta)-\beta(1-\frac{1}{2}x)~, 
\label{FadinKuraevBetaOrder}       
\end{equation}

The integration  over variable x  gives the following expression
\begin{equation}
\sigma(W)=\frac{3\pi \Gamma_{ee}\Gamma_h }{M^2 W \Gamma}  \bigg[
(1+\delta) \frac{4 \pi \beta }{\sin {\pi\beta}} \Imag{f}  -  
              \beta \mathcal{D}(W) \bigg]~,
\end{equation}
where 
\begin{equation}
\begin{split}
f=&\bigg(\frac{M}{W}+i\frac{\Gamma}{2 W} \bigg) \\
&\times \bigg(\frac{W}{M+W+i\frac{\Gamma}{2}}\bigg)^{1-\beta} \bigg(\frac{W}{M-W-i\frac{\Gamma}{2}}\bigg)^{1-\beta}
\end{split}
\end{equation} 
and 
\begin{equation}
 \begin{split}
\mathcal{D}(W) =&\frac{2M}{W}\Bigg(1+\bigg(\frac{M}{W}\bigg)^2\Bigg)  
\Bigg(\arctan{\frac{2M}{\Gamma}} - \arctan{\frac{2(M-W)}{\Gamma}}\Bigg) \\
-& \frac{\Gamma}{2 W}\bigg(1+3\bigg(\frac{M}{W}\bigg)^2\bigg)
\ln\frac{M^2+\frac{\Gamma^2}{4}}{(M-W)^2+\frac{\Gamma^2}{4}} \\
+&\frac{3M \Gamma}{W^2}\Bigg[1-\frac{\Gamma}{2W}\bigg(\arctan{\frac{2M}{\Gamma}}
- \arctan{\frac{2(M-W)}{\Gamma}}\bigg)\Bigg] \\
+&\frac{\Gamma}{2 W}-\frac{M\Gamma}{W^2}+\frac{\Gamma^3}{8{W}^3}
\ln\frac{M^2+\frac{\Gamma^2}{4}}{(M-W)^2+\frac{\Gamma^2}{4}}~.
\end{split}
\label{BWnorelativistic}
\end{equation}

For relativistic Breit-Wigner form given by expression:
\begin{equation}
\sigma(s)=\int  \!\,\frac{12\pi\Gamma_{ee}\Gamma_h}{ (s(1-x)-M^2)^2+\Gamma^2 M^2}\mathcal{F}(x,s) dx~
\end{equation}
where $s=W^2$ ; the integration over the variable x gives 
\begin{equation}
  \begin{split}
    \sigma(s) =&\frac{12\pi\Gamma_{ee} \Gamma_{h}} {\Gamma M s }
    \Bigg[
    (1+\delta)\frac{\pi\beta}{\sin{\pi\beta}} \Imag{f_{r}}  \\
    -&\frac{\beta}{2}\bigg(1+\frac{M^2}{s}\bigg)\bigg(\arctan{\frac{M}{\Gamma}}-
    \arctan{\frac{M^2-s}{\Gamma M}} \bigg) \\
    +&\frac{\beta \Gamma M}{4 s}
    \ln{\frac{\big(\frac{M^2}{s}\big)^2+\big(\frac{\Gamma
          M}{s}\big)^2}{\big(1-\frac{M^2}{s}\big)^2+\big(\frac{\Gamma
          M}{s}\big)^2}} \Bigg]~,
\end{split}
\label{BWrelativistic}
\end{equation}
where 
\begin{equation}
f_{r} = \Bigg(\frac{M^2}{s}-1-i \frac{\Gamma M}{s} \Bigg)^{\beta-1}~.
\label{frelativistic}
\end{equation}

In the case, where both resonant and non-resonant parts are present
in the decay amplitude, the expression for cross section reads
\begin{equation}
\sigma(s) =\int  \!\, \Big|\frac{\sqrt{12\pi\Gamma_{ee}\Gamma_{h}}}
{s(1-x)-M^2+i\Gamma M} + C e^{i\phi} \Big|^2 \mathcal{F}(x,s) dx~, 
\end{equation}
where C is constant,  and $\phi$ is phase angle of the interference.
After integrating over x  the cross section over can be represented
in the form 
\begin{equation}
\begin{split}
\sigma(s)=&\sigma_{R}(s)+ \sigma_{NR}(s) \\
 -& 2\frac{\sqrt{12\pi\Gamma_{ee}\Gamma_{h}}}{s}
 \frac{C(1+\delta)\pi\beta}{\sin{\pi\beta}} \Real{(f_{r}
   e^{-i\phi})}~,
\label{BWIntC}
\end{split}
\end{equation}
where $\sigma_{R}(s)$,$\sigma_{NR}(s)$ are resonant 
and non-resonant parts of the cross section, $f_{r}$ is the function 
introduced in (\ref{frelativistic}),
and the third term corresponds to the interference 
between the resonant and non-resonant hadron production.

Let us consider the interference between two resonances with 
masses $M_{1} $ and $M_{2}$, in that case, the cross section is :
\begin{equation}
\begin{split}
\sigma(s) & = \int  \!\, \Bigg|
\frac{\sqrt{12\pi\Gamma_{1,ee}\Gamma_{1,h}}}{s(1-x)-M_{1}^{2}+i\Gamma_{1}  M_{1}} \\
        +&
        \frac{\sqrt{12\pi\Gamma_{2,ee}\Gamma_{2,h}}}{s(1-x)-M_{2}^{2}+i\Gamma_{2} M_{2}} e^{i\phi} \Bigg|^2 \mathcal{F}(x,s) dx~, 
\end{split}
\end{equation}
where
$\Gamma_{1},\Gamma_{1,h},\Gamma_{1,ee},\Gamma_{2},\Gamma_{2,h},\Gamma_{2,ee}$
are total and partial widths of the resonances, and $\phi$ is phase angle of the interference.
The result of calculation is:
\begin{equation}
    \begin{split}
\sigma(s)=& \sigma_{1,R}(s)+ \sigma_{2,R}(s)  \\
 -& \frac{24 \pi \sqrt{\Gamma_{1,ee}\Gamma_{1,h} \Gamma_{2,ee}\Gamma_{2,h}} }{s}
  \frac{(1+\delta)\pi\beta}{\sin{\pi\beta}} \\
\times & \Real{ \Bigg(\frac{f_{1} e^{-i\phi}- f_{2} e^{i\phi}}{M_{1}^2-M_{2}^2-i(\Gamma_{1} M_{1}+ \Gamma_{2} M_{2} )} \Bigg)}~,
\label{BWInt12}
  \end{split}
\end{equation}
where  $\sigma_{1,R}(s)$, $\sigma_{2,R}(s)$  correspond to
resonances cross sections, and the last term represents  the
interference between resonances. 
The functions $f_{1}$ and $f_{2}$ are similar to function introduced in
$(\ref{frelativistic})$ :
\begin{equation}
f_{1}=\Bigg(\frac{M_{1}^2}{s}-1-i\frac{\Gamma_{1}  M_{1}}{s}\Bigg)^{\beta-1} ~,
\end{equation}
\begin{equation}
f_{2}=\Bigg(\frac{M_{2}^2}{s}-1-i\frac{\Gamma_{2}  M_{2}}{s}\Bigg)^{\beta-1} ~.
\end{equation}

It should be noted that although we have not  discussed the energy 
dependence of total width, our results are also applicable in such
case. If the production cross section has threshold behaviour our
approximation can be used if $\frac{\Gamma}{M-\sqrt{s_{thr}}}<<1$.

We also stress that equations
(\ref{BWnorelativistic}),(\ref{BWrelativistic}),
(\ref{BWIntC}),(\ref{BWInt12}) are applicable 
if  $\frac{\Gamma}{W} \sim \beta , \frac{\Gamma}{W}<<1 $.  

To summarize, the cross section given by integration 
over the probability  to emit certain fraction of energy from initial
state has been calculated for the  relativistic and nonrelativistic 
Breit-Wigner forms. 
 
The author would likes to thank A.G. Shamov for helpful discussions
and S.I. Eidelman for interest to this work. 


\begin{thebibliography}{3}
\bibitem{bib:FADIN}
E.\,A. Kuraev and V.\,S. Fadin, Yad. Fiz. {\bf 41 }, 733 (1985)

\bibitem{bib:JACKSON}
J.\,D. Jackson and D.\,D. Scharre, Nucl. Instrum. Methods. {\bf 128}, 13 (1975)

\bibitem{bib:ASIMOV}
Ya.\,I. Asimov,  A.\,I. Vainshtein, L.\,N. Lipatov et al., JETP Lett. {\bf 21}, 172 (1975)

\end{thebibliography}
\end{document}